\documentclass[
  prd, notitlepage, longbibliography, twocolumn,preprintnumbers, nofootinbib
  ]{revtex4-2}
\usepackage[utf8]{inputenc}

\usepackage{bm}
\usepackage{comment} 
\usepackage[colorlinks=true,urlcolor=blue,anchorcolor=black,citecolor=blue,linkcolor=black,filecolor=black,menucolor=black,pagecolor=black,linktocpage=true,pdfproducer=medialab,pdfa=true]{hyperref}
\usepackage{graphicx}
\usepackage{amsmath,latexsym,amssymb,mathrsfs,ascmac}
\usepackage{multirow}
\usepackage{braket}

\begin{document}

\preprint{KOBE-COSMO-22-06}

\title{Implications of the singularity theorem for the size of a nonsingular universe}

\author{Kimihiro Nomura}
\email{knomura@stu.kobe-u.ac.jp}
\affiliation{Department of Physics, Kobe University, Kobe 657-8501, Japan}
 
\author{Daisuke Yoshida}
\email{dyoshida@math.nagoya-u.ac.jp}
\affiliation{Department of Mathematics, Nagoya University, Nagoya 464-8602, Japan}

\begin{abstract}
A general property of universes without initial singularity is investigated based on the singularity theorem, assuming the null convergence condition and the global hyperbolicity.
As a direct consequence of the singularity theorem, the universal covering of a Cauchy surface of a nonsingular universe with a past trapped surface must have the topology of $S^3$.
In addition, we find that the affine size of a nonsingular universe, defined through the affine length of null geodesics, is bounded above. In the case where a part of the nonsingular spacetime is described by Friedmann-Lema\^itre-Robertson-Walker spacetime, we find that this upper bound can be understood as the affine size of the corresponding closed de Sitter universe. We also evaluate the upper bound of the affine size of our Universe based on the trapped surface confirmed by recent observations of baryon acoustic oscillations, assuming that our Universe has no initial singularity.  
\end{abstract}

\maketitle
\section{Introduction}
The singularity theorems \cite{Penrose:1964wq,Hawking:1967ju,Hawking:1970zqf,Tipler:1978zz} state that a spacetime singularity is formed in very common situations. The presence of singularity is regarded as an implication of the breakdown of general relativity. One reason for this is that typical examples of singularities, such as black hole singularities and the big bang singularity, involve the divergence of local curvature invariants, which is expected to induce the effect of unknown high energy physics such as quantum gravity. To know what actually happens in the real world in the situation where a singularity is predicted by general relativity, we need to understand physics in a deep ultraviolet (UV) region.

Several possibilities to resolve a singularity in a deep UV region may be considered. One possibility is that the concept of spacetime is applicable only in an infrared region and it must be abandoned in a deep UV region. However, in order to advance our understanding along this scenario, we need to know more about quantum gravity itself. On the contrary, in this paper, we will focus on alternative scenarios which can be approached in a bottom-up manner without referring to the details of quantum gravity. We assume that the concept of spacetime will not be broken near the singularity but some corrections to the Einstein equations, which may originate from UV physics, will prevent the production of the singularity. As a result, the singularity is resolved within a (semi)classical treatment. Thus, it must be important to study what kind of classical gravitational theory can resolve a singularity. Toward the gravitational theory without singularity, a limiting curvature mechanism has been studied both in the context of cosmological \cite{Mukhanov:1991zn,Brandenberger:1993ef,Moessner:1994jm,Easson:2006jd,Yoshida:2017swb,Quintin:2019orx,Sakakihara:2020rdy, Frolov:2021ayq} and black hole singularities \cite{Trodden:1993dm,Easson:2002tg,Chamseddine:2016ktu,Easson:2017pfe,Yoshida:2018kwy,Frolov:2021afd}. An example of UV theories to realize the limiting curvature mechanism is loop quantum cosmology \cite{Ashtekar:2006uz}, where quantum corrected Friedmann equations prohibit the divergence of the Hubble parameter.

If the classical picture of spacetime is valid near the singularity, we can approach unknown UV physics by investigating a property of nonsingular spacetimes directly, without referring to the dynamics of gravity in detail. In the context of black hole singularity, for example, Bardeen proposed a possible geometry of a regular black hole that does not have singularity but looks like a Schwarzschild black hole from observers in the asymptotic region \cite{bardeen1968non}. 
Since then, various regular black hole geometries have been proposed and studied \cite{Ayon-Beato:1999kuh,Dymnikova:2004zc, Hayward:2005gi, Ayon-Beato:2004ywd,Fan:2016hvf,Maeda:2021jdc}.
One important property of Bardeen's black hole is that it is not globally hyperbolic and has compact time slices inside the black hole. 
In Ref.~\cite{Borde:1996df}, Borde showed that this property holds for a more general class of regular black holes based on the singularity theorem.

The purpose of this paper is to investigate a general property of nonsingular universes from the viewpoint of the singularity theorems by Penrose \cite{Penrose:1964wq} and Tipler \cite{Tipler:1978zz}, inspired by Borde's research for regular black holes.
An important point here is that the singularity theorem is a theorem on the geometry of a spacetime and hence it is applicable even when the dynamics of gravity is not described by the Einstein equations, for example, even when the corrections from unknown UV physics are relevant. In this paper, we will investigate a general property of nonsingular universes that are consistent with the null convergence condition.
The simplest example of such spacetimes is closed de Sitter space. It is also known that the universe remains nonsingular if sufficiently small anisotropies are added to the closed de Sitter universe \cite{Bramberger:2019zez}. Even for flat Friedmann-Lema\^itre-Robertson-Walker (FLRW) geometry, one can construct a nonsingular universe by considering the maximal extension of an inflationary universe as studied in Refs.~\cite{Yoshida:2018ndv, Nomura:2021lzz, Nishii:2021ylb}.
What we will show in this paper is that any such nonsingular universe has a compact Cauchy surface and the size of the universe, defined later through the affine length of null geodesics, is smaller than the exact de Sitter space.

This paper is organized as follows.
In the next section, we review the singularity theorem by Penrose and clarify the implications for nonsingular spacetimes.
There we will see that a Cauchy surface of a nonsingular universe
that possesses a trapped surface and satisfies
the null convergence condition must be compact, especially the topology of the universal covering of a Cauchy surface must be $S^{3}$. In addition, we will see that the size of the universe measured by the affine length of null geodesics has an upper bound.  
Then, in Sec.~\ref{sec:FLRW}, we focus on homogeneous and isotropic spacetimes and find that the size of any nonsingular homogeneous and isotropic universe is smaller than the corresponding de Sitter universe. In Sec.~\ref{sec:example}, we study explicit examples of nonsingular homogeneous and isotropic spacetimes and confirm the size of the universe is smaller than de Sitter space.
Then we apply our result to our realistic universe in Sec.~\ref{sec:observation}.
The final section is devoted to the summary and discussions.

\section{Singularity theorem and nonsingular universes}
\label{sec:singularitytheorem}
\subsection{Singularity theorem}

We start by reviewing the singularity theorem by Penrose \cite{Penrose:1964wq} and its extension by Tipler \cite{Tipler:1978zz}.
We are interested in the initial singularity of an expanding universe, whereas the original Penrose theorem is concerned with singularities that are formed in the future, like a black hole spacetime. After flipping future to past, Penrose's first singularity theorem can be rephrased as follows.
\\
\\
{\it
For a spacetime $({\cal M}, g)$, the following four statements lead to a contradiction:\\
(i) There is a noncompact Cauchy surface ${\cal C}$ in ${\cal M}$.\\
(ii) The null convergence condition is satisfied.\\
(iii) There is a closed past trapped surface. \\
(iv) $({\cal M}, g)$ is past null geodesically complete.}
\\
\\
Here we quickly give definitions and interpretations of each condition.  For a more detailed description, we refer the reader to Refs.~\cite{Hawking:1973uf, Wald:1984rg}. On condition (i),  a Cauchy surface is defined as an achronal set ${\cal C}$ where all inextendible (i.e., already extended as much as possible) causal curves intersect with it. The spacetime with a Cauchy surface is said to be globally hyperbolic. 
Thus, assumption (i) can be rephrased as (i-i) the spacetime $({\cal M}, g)$ is globally hyperbolic, and (i-ii) a Cauchy surface is noncompact. In Ref.~\cite{Tipler:1978zz}, Tipler showed that Penrose's theorem holds if the topology of the universal covering of the Cauchy surface is not $S^3$.  Therefore, the assumption (i-ii) can be weakened as (i-ii)$'$ the topology of the universal covering of a Cauchy surface is not $S^3$.
See also Ref.~\cite{Galloway:2017mts} for another discussion based on the prime decomposition theorem for three-manifolds.
On the condition (ii), the null convergence condition is defined by $R_{ab} k^{a} k^{b} \geq 0$ for all null vectors $k^{a}$, where $R_{ab}$ is the Ricci tensor.
When the Einstein equations hold,
the null convergence condition reduces to the null energy condition for the energy-momentum tensor $T_{ab}$, $T_{ab} k^{a} k^{b} \geq 0$.
On condition (iii), a closed two-dimensional surface ${\cal T}$ is said to be a closed past trapped surface if all the past-directed null geodesic congruences that are orthogonal to ${\cal T}$ have a negative expansion; in other words, all the future-directed null geodesic congruences that are orthogonal to ${\cal T}$ have a positive expansion. In this paper, we take an affine parameter of geodesics to be increasing toward the future directions. 
Then the expansion of a null geodesic congruence is defined as follows.  Let $k$ be the tangent vectors of affine parametrized null geodesics and $l$ be a null vector field which satisfies $l_{\mu} k^{\mu} = -1$. Then we can define an induced metric $\widehat{h}_{\mu\nu}$ by
\begin{align}
 \widehat{h}_{\mu\nu} = g_{\mu\nu} + l_{\mu} k_{\nu} + k_{\mu} l_{\nu}.
\end{align}
Then the tensor $\widehat{B}_{\mu\nu}$ can be defined by
\begin{align}
\widehat{B}_{\mu\nu} = \widehat{h}_{\mu}{}^{\rho} \widehat{h}_{\nu}{}^{\sigma} \nabla_{\rho} k_{\sigma},
\end{align}
and the expansion is defined as the trace of the tensor $\widehat{B}_{\mu\nu}$,
\begin{align}
 \theta = \widehat{h}^{\mu\nu} \widehat{B}_{\mu\nu}. 
\end{align}
Note that the $\theta$ does not depend on the choice of $l_{\mu}$.
Since the affine parameter is taken to be increasing toward the future directions, this $\theta$ measures the expansion of the future-directed null geodesic congruences. Thus, the condition for ${\cal T}$ to be a past trapped surface is that $\theta$ is everywhere positive on ${\cal T}$ for all the null geodesic congruences orthogonal to ${\cal T}$.

Usually, the singularity theorem is understood to predict a singularity. For a spacetime satisfying assumptions (i) - (iii), the other assumption (iv) cannot be satisfied; that is, the spacetime must have an incomplete null geodesic.  In this paper, contrarily, we will focus on a {\it nonsingular} universe. Assuming (i-i), (ii), (iii), and (iv), the singularity theorem states that the assumption (i-ii)$'$ cannot be satisfied; that is, the topology of the universal covering of a Cauchy surface must be $S^3$.
Thus, the singularity theorem can be rephrased as follows.
\\
\\
{\it
Suppose that a spacetime $({\cal M}, g)$ satisfies the following conditions:\\
(i-i) $({\cal M}, g)$ is globally hyperbolic.\\
(ii) The null convergence condition is satisfied.\\
(iii) There is a closed past trapped surface. \\
(iv) $({\cal M}, g)$ is past null geodesically complete.\\
Then the topology of the universal covering of a Cauchy surface is $S^3$.
}
\\
\\
The de Sitter space is an example that satisfies all the assumptions and has a Cauchy surface with the topology of $S^3$. In contrast, the flat de Sitter universe on three-torus satisfies assumptions (i-i), (ii), and (iii), but has a Cauchy surface with the topology of $T^3$. Hence, it is past incomplete \cite{Galloway:2004bk} and actually has a quasiregular singularity \cite{Numasawa:2019juw}. We would like to emphasize that this does not mean the necessity of the spatial curvature for a FLRW universe to be geodesically complete. As we will see in an example later, the maximal extension of a flat FLRW universe could have time surfaces with the topology of $S^{3}$ and could be geodesically complete, as in the case of de Sitter space. Note that the same results are obtained also based on the prime decomposition theorem for three-manifolds in Ref.~\cite{Galloway:2017mts}.

\subsection{Bound on the size of a nonsingular universe}
\label{sec:2B}
Here we would like to highlight another aspect of the singularity theorem. The proof of the singularity theorem has information on the size of a nonsingular universe. Let us see this in detail.

\begin{figure}[htbp]
 \begin{center}
  \includegraphics[width=\hsize]{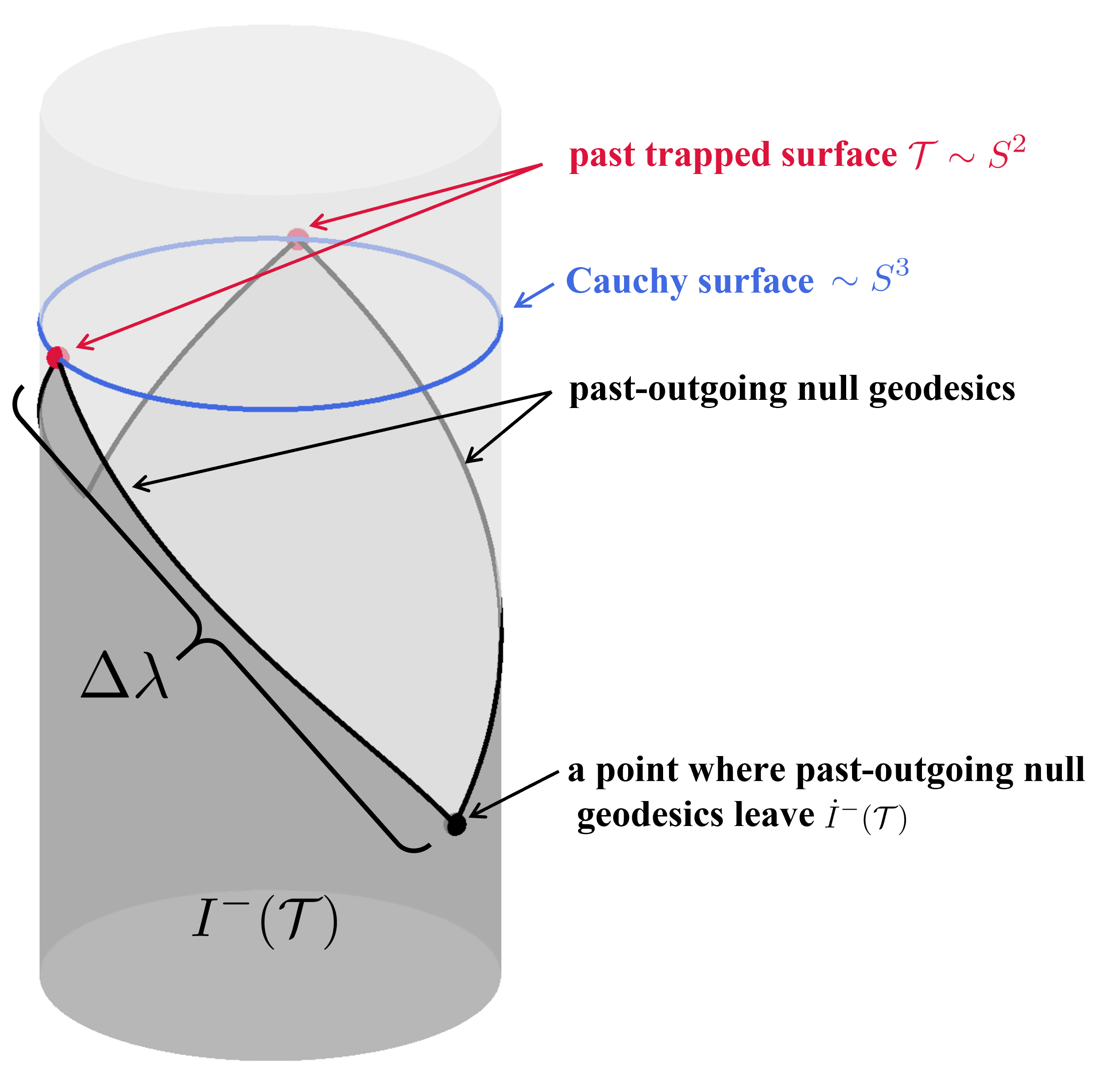}
\caption{A rough sketch of the definition of $\Delta \lambda$:
The entire cylinder represents a 1+1-dimensional part of a 1+3-dimensional spacetime with a Cauchy surface $\sim S^3$ (the blue circle).
The past trapped surface ${\cal T} \sim S^2$ is represented by the red points in the 1+1-dimensional part. 
From each point, one can draw past-ingoing and past-outgoing null geodesics, which generate the boundary of the chronological past $\dot{I}^{-}(\cal T)$.
The past-outgoing null geodesics leave $\dot{I}^{-}(\cal T)$ and enter the interior of $I^{-}({\cal T})$ at the black point.
Then the outward affine size of the universe $\Delta \lambda$ is defined as the maximal value of affine length between the red point and the black point.   
}
\label{fig1}
 \end{center}
\end{figure}

Let us consider a spacetime that satisfies the above assumptions (i-i), (ii), (iii), and (iv). 
For a given trapped surface ${\cal T}$, we define {\it the outward affine size of the universe $\Delta \lambda$ from the trapped surface ${\cal T}$} by the maximum affine length of the past-outgoing null geodesics orthogonal to ${\cal T}$, which have end points on ${\cal T}$ and $\dot{I}^{-}({\cal T})$, where $\dot{I}^-({\cal T})$ denotes the boundary of the chronological past of ${\cal T}$ and we assume that the ambiguity of the affine parameter is fixed on ${\cal T}$.  
Since every past-outgoing null geodesic with an affine length greater than $\Delta \lambda$ departs from $\dot{I}^{-}(\cal T)$ by the definition, the $\Delta \lambda$ characterizes the size of the universe measured by null geodesics (Fig.~\ref{fig1}). 

An important point here is that {\it the outward affine size of the universe $\Delta \lambda$ must be smaller than $2/\theta_{0}$, where $\theta_{0}$ is the minimum value of the expansion on ${\cal T}$}. We will show it as follows. 

Let $\{\gamma_{p}\}_{p \in {\cal T}}$ be a past-outgoing inextendible null geodesic congruence orthonormal to ${\cal T}$ with an affine parameter $\lambda$.  We fix the initial value of the affine parameter $\lambda_0$ by $\gamma_{p}(\lambda=\lambda_0) = p$. Then, the Raychaudhuri equation for twist-free null geodesics can be written as
\begin{align}
 & \frac{d \theta}{d \lambda} + \frac{1}{2} \theta^2 =  - \widehat{\sigma}_{\mu\nu} \widehat{\sigma}^{\mu\nu}
 - R_{\mu\nu}k^{\mu} k^{\nu} \leq 0,
\end{align}
where the inequality is obtained from the null convergence condition $R_{\mu\nu}k^{\mu} k^{\nu} \geq 0$. Here $\widehat{\sigma}_{\mu\nu}$ is the shear tensor defined as the trace-free and symmetric part of $\widehat{B}_{\mu\nu}$. By integrating this equation, we obtain $ \theta^{-1}(\lambda) \leq \theta^{-1}_0 + \frac{1}{2} (\lambda - \lambda_0)$ for $\lambda  < \lambda_0$. By the definition of the past trapped surface, $\theta_{0}^{-1}$ must be positive. Since the right-hand side is negative for $\lambda - \lambda_0 < -2\theta_0^{-1}$, there is the affine parameter $\lambda_{c}(< \lambda_0)$ that satisfies $\theta^{-1}(\lambda_{c}) = 0 $ and $0 > \lambda_c - \lambda_0 \geq - 2 \theta_0^{-1}$. A point $\gamma_{p}(\lambda_{c})$ is called a conjugate point. Thus, from assumptions (ii) - (iv), we can say that all the past-inextendible null geodesics starting from and orthogonal to the trapped surface ${\cal T}$ have a conjugate point at an affine parameter $\lambda_{c}$, which satisfies $\lambda_c - \lambda_0 \geq - 2 \theta_0^{-1}$. On the other hand, the following theorem (Wald \cite{Wald:1984rg}, Theorem 9.3.11, with replacing future with past) is satisfied in a globally hyperbolic spacetime:
\\
\\
{\it
Let $({\cal M}, g)$ be a globally hyperbolic spacetime and let $K$ be a compact, orientable, two-dimensional spacelike submanifold of ${\cal M}$. Then every $p \in \dot{I}^{-}(K)$ lies on a past-directed null geodesic starting from $K$, which is orthogonal to $K$ and has no point conjugate to $K$ between $K$ and $p$.  
}
\\
\\
Thus, from assumption (i-i), all the past-inextendible null geodesics must depart from $\dot{I}^{-}({\cal T})$ at an affine parameter $\lambda_{*}(< \lambda_0)$, which satisfies $0 > \lambda_{*} - \lambda_0 \geq \lambda_{c} - \lambda_0 \geq - 2\theta_{0}^{-1}$.
Thus, the above inequality can be understood as that the outward affine size of the universe $\Delta \lambda = |\lambda_* - \lambda_0|$ is bounded by the expansion of the current universe $\theta_{0}$ by $\Delta \lambda \leq 2/ \theta_{0}$.

In principle, the outward affine size $\Delta \lambda$ defined above depends on the normalization of the affine parameter (scaling of the null vector). Note, however, that the inequality $\Delta \lambda \leq 2/ \theta_{0}$ always holds independent of the normalization since both sides scale the same way under the change of the normalization. Of course, once the normalization is specified appropriately on the trapped surface, the size $\Delta \lambda$ completely makes sense.

\section{Nonsingular FLRW universes must be smaller than de Sitter space}
\label{sec:FLRW}
The discussion in the previous section can be applied to any nonsingular spacetime.
In the following, we particularly focus on a spacetime $({\cal M}, g)$ that includes a homogeneous and isotropic expanding universe as a subregion ${\cal M}_{\rm FLRW} \subset {\cal M}$, where the metric can be expressed as
\begin{align}
 g_{\mu\nu}dx^{\mu} dx^{\nu} = - dt^2 + a(t)^2 \left(d\chi^2 + \Phi_{k}(\chi)^2 d\Omega^2 \right),
\end{align}
where the scale factor $a$ is an increasing function of time, $\partial_t a > 0$,
$\Phi_{k}$ is given by
\begin{align}
 \Phi_{k}(\chi) = 
\begin{cases}
 \sin \chi &( k = +1)\\
 \chi&(k = 0) \\ 
 \sinh \chi& ( k = -1)
\end{cases},
\end{align}
depending on the sign of the spatial curvature $k$, and 
$d\Omega^2$ is the metric of the two-sphere which can be described by the polar coordinates $(\theta, \phi)$ as
\begin{align}
  d\Omega^2 
  = d\theta^2 + \sin^2 \theta d\phi^2.
\end{align}
Note that the coordinate $\chi$ is defined in $(0, \infty)$ for $k = 0, -1$, while in $(0, \pi)$ for $k = +1$.

\subsection{Past trapped surface in a FLRW universe}
Let us consider a future-ingoing (in other words, past-outgoing) null geodesic congruence orthogonal to a two-sphere, say ${\cal T}$, defined by $t = t_{0}$ and $\chi = \chi_{0}$.
We characterize each geodesic by the polar coordinates of the point where the geodesic intersect with ${\cal T}$, as $\gamma_{\theta, \phi}(\lambda)$. We set the initial value of the affine parameter $\lambda_0$ by $\gamma_{\theta, \phi}(\lambda = \lambda_0) \in {\cal T}$.  
The tangent vector ${k}(\theta, \phi)$ of a null geodesic $\gamma_{\theta,\phi}(\lambda)$ can be expressed as
\begin{align}
 {k} = \frac{1}{a(t)} \boldsymbol{\partial}_{t} - \frac{1}{a(t)^2} \boldsymbol{\partial}_{\chi},
  \label{tangent}
\end{align}
where the coordinate bases are denoted by $\boldsymbol{\partial}_\mu$. 
Of course, there is an ambiguity in the definition of the affine parameter of each geodesic. 
In general, one can introduce an overall factor of an arbitrary function of $\theta$ and $\phi$ in Eq.~\eqref{tangent}. 
The definition of the affine parameter that we use here can be characterized by
\begin{align}
 d \lambda = a(t) d t.
\end{align}
The induced metric $\widehat{h}_{\mu\nu}$ and the tensor $\widehat{B}_{\mu\nu}$ for such a null geodesic congruence can be evaluated as
\begin{align}
\widehat{h}_{\mu\nu} dx^{\mu} dx^{\nu} &= a(t)^2 \Phi_{k}(\chi)^2 d \Omega^2, \\
\widehat{B}_{\mu\nu} dx^{\mu} dx^{\nu} &= \Phi_{k}(\chi)^2 \left( a(t) H(t) - \frac{\Phi_{k}'(\chi)}{\Phi_{k}(\chi)}\right)  d\Omega^2.
\end{align}
Here we use ${l} =  (a \boldsymbol{\partial}_{t} +  \boldsymbol{\partial}_{\chi})/2$ and introduce the Hubble parameter $H(t) = \partial_t a / a$.
Then the expansion of this congruence on ${\cal T}$ can be evaluated as 
\begin{align}
 \theta_{0} (\theta, \phi) &= \widehat{h}^{\mu\nu} \widehat{B}_{\mu\nu}|_{\cal T} \notag\\
&= \frac{2}{a(t_{0})^2} \left( a(t_{0}) H(t_{0}) -  \frac{\Phi_{k}'(\chi_{0})}{\Phi_{k}(\chi_{0})}\right).
\end{align}
Therefore, the two-sphere ${\cal T}$ at $t=t_{0}, \chi = \chi_{0}$ is a trapped surface if the following inequality is satisfied:
\begin{align}
  a(t_{0}) H(t_{0}) >  \frac{\Phi_{k}'(\chi_{0})}{\Phi_{k}(\chi_{0})} = 
\begin{cases}
 1/\tan \chi_{0} & (k = +1)\\
 1/\chi_{0} &(k = 0)\\
 1/\tanh \chi_{0} & (k=-1)
\end{cases}.
\label{trapped}
\end{align}

Clearly, one can always find a past trapped surface in a closed ($k=+1$) or flat ($k=0$) FLRW universe by considering a two-sphere with a radius $\chi_{0}$ close to $\pi/2$ (for $k= +1$) or large enough (for $k=0$).
For the $k = -1$ case, a trapped surface exists only when $a(t_0) H(t_0) > 1$.
This condition is always satisfied if the universe follows the Friedmann equation with the weak energy condition for the energy density $\rho$, i.e., $\rho > 0$, because $a^2 H^2 - 1 = (8 \pi G/3) \rho a^2 > 0,$ with $G$ being the Newton constant.

To summarize, if a sufficiently large part of a universe is expressed by a FLRW universe,
so that there exists a sufficiently large two-sphere in it, a two-sphere ${\cal T}$ defined by $t = t_{0}, \chi = \chi_{0}$ is a past trapped surface.
Then, by the singularity theorem, assuming the global hyperbolicity, the null convergence condition, and the absence of initial singularity, 
we can say that the outward affine size of the universe $\Delta \lambda$ from the trapped surface ${\cal T}$ is smaller than $2/\theta_{0}$,
\begin{align}
 \Delta \lambda \leq \frac{2}{\theta_{0}} = \frac{ a(t_{0})^2}{a(t_{0}) H(t_{0}) - \frac{\Phi_{k}'(\chi_{0})}{\Phi_{k}(\chi_{0})}}.
 \label{lambdaineq}
\end{align}

\subsection{de Sitter space as the largest space}
\label{subsec:dSlargest}
In this subsection, we clarify an interpretation of the inequality \eqref{lambdaineq}, which means that the size of a FLRW universe is smaller than that of the (maximally extended) de Sitter universe.
To see this, let us focus on the closed de Sitter universe $(\tilde{\cal M}, \tilde{g})$, where the metric is given by 
\begin{align}
 \tilde{g}_{\mu\nu} dx^{\mu} dx^{\nu} = - dt^2 + \tilde{a}(t)^2 \left(d \chi^2 + \sin^2 \chi d\Omega^2 \right),
\end{align}
with a scale factor
\begin{align}
 \tilde{a}(t) = \frac{1}{H_{\rm dS}} \cosh\left( H_{\rm dS} (t - t_{\rm dS})\right).
\end{align}
Here $H_{\rm dS}$ and $t_{\rm dS}$ are free parameters that determine the de Sitter radius and the origin of time, respectively.
By introducing the conformal time $\eta$ by
\begin{align}
\eta = \text{Arctan} \left( \sinh (H_{\rm dS}(t - t_{\rm dS}))\right),
\end{align}
the metric can be written as
\begin{align}
 \tilde{g}_{\mu\nu} dx^{\mu} dx^{\nu} = \tilde{a}(\eta)^2 \left(- d\eta^2 + \left(d \chi^2 + \sin^2 \chi d\Omega^2 \right) \right).
\end{align}
Here the scale factor $\tilde{a}(\eta)$ can be represented as a function of $\eta$ by
\begin{align}
 \tilde{a}(\eta) = \frac{1}{H_{\rm dS}} \frac{1}{\cos \eta},
\end{align}
and hence the Hubble parameter can be evaluated as
\begin{align}
 \tilde{H}(\eta) = H_{\rm dS} \sin \eta.
\end{align}

Let us consider a two-sphere ${\cal T}$ characterized by $t = t_{0}$ and $\chi = \chi_{0}$ in the de Sitter space and the congruence of the past-outgoing null geodesics orthonormal to ${\cal T}$ characterized by $\eta(\lambda) = \eta_{0} - ( \chi(\lambda) - \chi_{0})$.
As discussed in the previous subsection, the expansion of such a congruence can be written as
\begin{align}
 \frac{2}{\tilde{\theta}_{0}} = \frac{\tilde{a}_{0}^2}{\tilde{a}_{0} \tilde{H}_{0} - \frac{1}{\tan \chi_{0}}} ,
\end{align}  
where $\tilde{a}_0 = \tilde{a}(\eta_0)$ and $\tilde{H}_0 = \tilde{H}(\eta_0)$.
By tracing each geodesic to the past, all the geodesics depart from $I^{-}({\cal T})$ after they arrive at the south pole of the $S^3$.
Hence, the outward affine size of the de Sitter space, say $\Delta \lambda^{\rm dS} = |\lambda_{*}^{\rm dS} - \lambda_0|$, can be obtained by $\chi(\lambda_{*}^{\rm dS}) = \pi$.
Since $d \lambda = \tilde{a}(\eta)^2 d \eta$, $\Delta \lambda^{\rm dS}$ can be evaluated as
\begin{align}
  \Delta \lambda^{\rm dS} &= \int_{\eta_{0} - (\pi - \chi_{0})}^{\eta_{0}} \frac{1}{H_{\rm dS}^2} \frac{d \eta}{\cos^2 \eta} \notag\\
&= \frac{1}{H_{\rm dS}^2} \left( \tan \eta_{0} - \tan (\eta_{0} - (\pi - \chi_{0}))\right) \notag\\
&= \frac{1}{H_{\rm dS}^2} \left( \tan \eta_{0} - \frac{\tan \eta_{0} - \tan( \pi - \chi_{0})}{1 + \tan \eta_{0} \tan(\pi - \chi_{0})}\right) .
\end{align}
Using the relation $H^2_{\rm dS} = (1 + \tilde{H}_{0}^2 \tilde{a}_{0}^2)/\tilde{a}^2_{0}$ and $\tan \eta_{0} = \tilde{a}_{0} \tilde{H}_{0}$,
 the expression can be simplified as
\begin{align}
 \Delta \lambda^{\rm dS} = \frac{\tilde{a}_{0}^2}{\tilde{a}_{0} \tilde{H}_{0} - \frac{1}{\tan \chi_{0}}} = \frac{2}{\tilde{\theta}_{0}}.
\end{align}
Thus the de Sitter universe saturates the inequality $\Delta \lambda^{\rm dS} \leq 2/\tilde{\theta}_{0}$.
We would like to emphasize that the affine length and the value of the expansion are coordinate independent quantities and hence the same relation holds if one describes the de Sitter universe in the different coordinates.

To see a physical interpretation of this result, let $({\cal M}, g)$ be a nonsingular FLRW universe (with or without the spatial curvature) that satisfies the null convergence condition and global hyperbolicity, and let us consider a two-sphere at $t = t_{0}$ and $\chi = \chi_{0}$ large enough to be a past trapped surface.
Then, from the general discussion in the previous section, the outward affine size of the universe has an upper bound: $\Delta \lambda < 2/\theta_{0}$. 
On the one hand, let $(\tilde{\cal M}, g)$ be a spacetime where the spacetime region of $t \leq t_{0}$ in $({\cal M}, g)$ is replaced by the corresponding de Sitter space that has the same value of the scale factor and Hubble parameter at $t = t_{0}$,
$a_0 = a(t_{0}) = \tilde{a}(t_{0})$ and $H_0 = H(t_{0}) = \tilde{H}(t_{0})$, as well as the same sign of the spatial curvature.
Explicitly, the scale factor of the de Sitter universe can be obtained as
\begin{align}
 a_{\rm dS}(t) = 
\begin{cases}
 \frac{1}{H_{\rm dS}} \cosh \left( H_{\rm dS}(t - t_{\rm dS}) \right) & ( k = +1),\\
 a_{0} \mathrm{e}^{H_{0}(t - t_{0})} & (k = 0),\\
 \frac{1}{H_{\rm dS}} \sinh \left( H_{\rm dS}(t - t_{\rm dS}) \right) & (k = -1),
\end{cases} 
\end{align} 
with
\begin{align}
 H_{\rm dS} &= \frac{\sqrt{k + a_{0}^2 H_{0}^2}}{a_{0}} ,
\end{align}
and
\begin{align}
  t_{\rm dS} &= 
\begin{cases}
 t_{0} - \frac{a_{0}}{\sqrt{1 + a_{0}^2 H_{0}^2}} \text{Arcsinh}(a_{0} H_{0}) & (k = +1)\\
 t_{0} - \frac{a_{0}}{\sqrt{- 1 + a_{0}^2 H_{0}^2}} \text{Arccosh}(a_{0} H_{0}) & (k= -1)
\end{cases}.
\end{align}
Since the expansions of two spacetimes coincide $2/\theta_0 = 2/\tilde{\theta}_0$, the general inequality $\Delta \lambda \leq 2/\theta_0$ can be written as 
\begin{align}
 \Delta \lambda \leq \Delta \lambda^{\rm dS}.
\end{align}
Thus, the affine size of any nonsingular FLRW universe is always smaller than that of a universe with the past region $t < t_{0}$ replaced with the de Sitter universe.
This is the main claim of this paper. 

\section{Examples}
\label{sec:example}

In the previous sections, we show that the outward affine size of any nonsingular universe from a past trapped surface has an upper bound in terms of the expansion of a null geodesic congruence on the trapped surface, under some assumptions. In this section, we propose two toy models of nonsingular universes and confirm that the above statement actually holds by explicitly evaluating the outward affine size and the expansion on a trapped surface and comparing them. 

\subsection{Scaled closed de Sitter universe}
As a simple example, let us consider a scaled closed de Sitter universe with the metric
\begin{align}
 g_{\mu\nu} dx^{\mu} dx^{\nu} = - dt^2 + \frac{\gamma^2}{ \bar{H}^2} \cosh^2 \bar{H} t \left( d \chi^2 + \sin^2 \chi d \Omega^2 \right),
 \notag \\
 (0 < \gamma \leq 1)
 \label{scaleddS}
\end{align}
where $\bar{H}$ is a positive constant, and the parameter $\gamma$ represents the relative size to the exact de Sitter space. In particular, the case of $\gamma = 1$ corresponds to the exact de Sitter space.
For a closed FLRW universe, the null convergence condition is equivalent to
the condition
\begin{align}
 \partial_{t} H - \frac{1}{a^2}  \leq 0.
\end{align}
Then one can check that our scale factor satisfies
\begin{align}
 \partial_{t} H - \frac{1}{a^2} = \frac{\gamma^2 - 1}{\gamma^2} \frac{\bar{H}^2}{\cosh^2 \bar{H} t}
\end{align}
and hence the scaled closed de Sitter universe is consistent with the null convergence condition for $\gamma \leq 1$.

\begin{figure}[htbp]
 \begin{center}
  \includegraphics[width=\hsize]{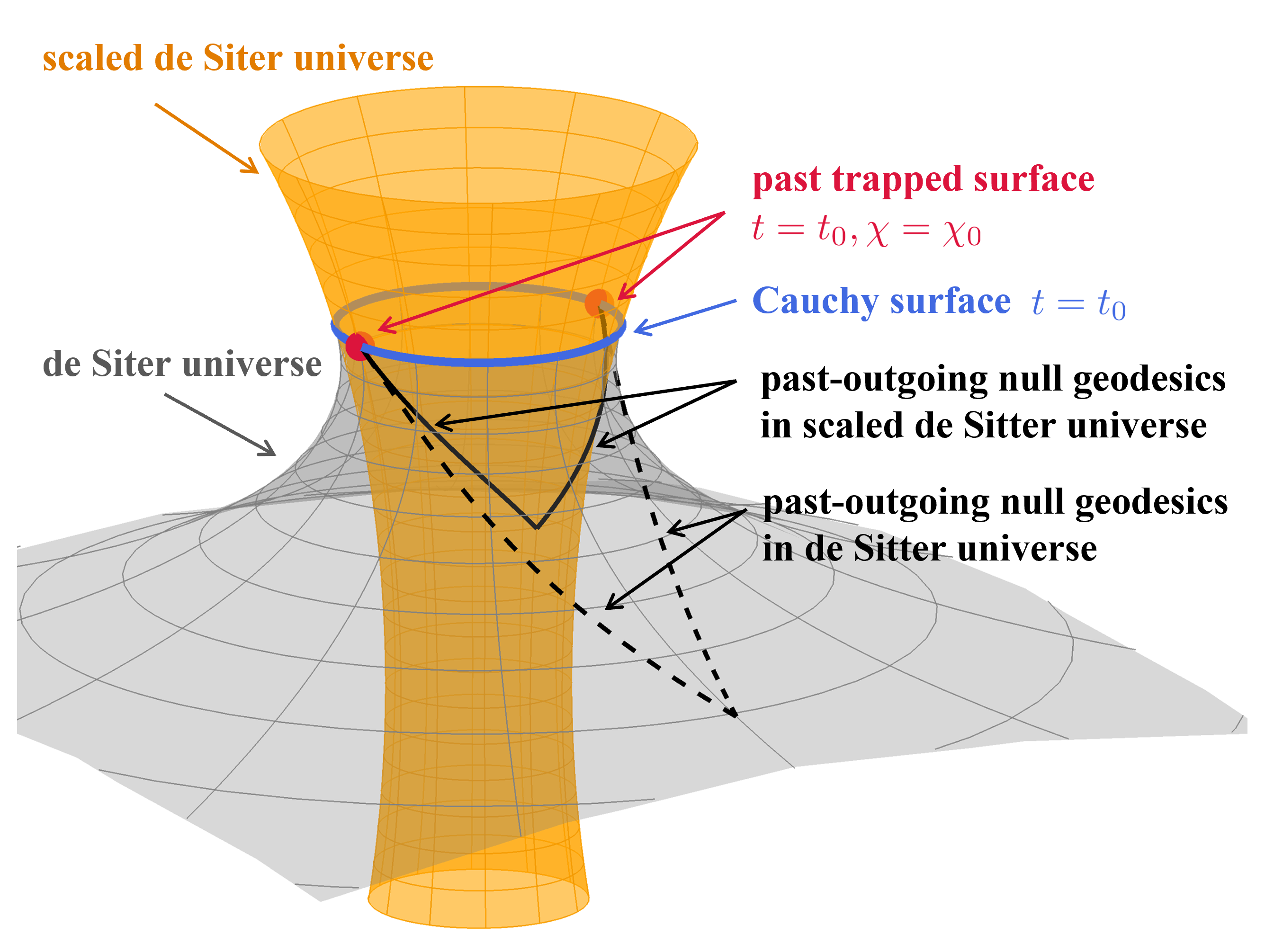}
\caption{
The de Sitter universe and scaled de Sitter universe are represented as hyperboloids embedded in Minkowski spacetime. The original scaled de Sitter universe \eqref{scaleddS} is the orange hyperboloid and the corresponding de Sitter universe attached to the past of $t = t_{0}$ is the gray one. The two universes have the same scale factor and Hubble parameter on $t = t_{0}$ time surface (blue curve).  A trapped surface at $t = t_{0}, \chi = \chi_{0} \sim \pi/2$ is described by the red points.  We claim that the affine length of the past-outgoing null geodesic starting from the trapped surface in the original scaled de Sitter universe (black, solid curve) is smaller than that of the corresponding de Sitter universe (black, dashed curve).}
\label{fig2}
 \end{center}
\end{figure}

For a given two-sphere ${\cal T}$ on a time slice $t = t_{0} > 0$ and an angle $\chi_{0}$, the expansion of the past-outgoing (thus future-ingoing) null geodesic congruence orthonormal to ${\cal T}$ can be evaluated as
\begin{align}
 \frac{2}{\theta_{0}} =  \frac{\gamma^2 \cosh^2 \bar{H} t_{0}}{\bar{H}^2} \frac{\tan \chi_{0}}{ \gamma \tan \chi_{0} \sinh \bar{H} t_{0} - 1}.
\end{align}
We are interested in the case in which ${\cal T}$ is a past trapped surface for past-outgoing congruences, that is, $\theta_0 > 0$. Thus, we assume $\chi_{0}$ satisfies\footnote{Note that, if $\chi_0$ is too close to $\pi$ (the south pole), the two-sphere there would not be a past trapped surface for {\it both} of the past-outgoing and past-ingoing null geodesic congruences since the expansion of a future-outgoing (i.e., past-ingoing) congruence would have a negative expansion. The purpose of this section is to compare the outward affine size measured by the past-outgoing geodesic with the expansion of the past-outgoing congruence on the two-sphere. Thus, here we focus only on the past-outgoing geodesics and do not care about the past-ingoing geodesics.}
\begin{align}
 \tan \chi_{0} > \frac{1}{\gamma \sinh \bar{H} t_{0}} \qquad \text{or} \qquad  \frac{\pi}{2} < \chi_{0} < \pi .
\end{align}
From the general discussion above, the value of $2/\theta_{0}$ corresponds to the outward affine size of the corresponding de Sitter space with a parameter
\begin{align}
 H_{\rm dS}& = \bar{H} \frac{\sqrt{1 + \gamma^2 \sinh^2 \bar{H} t_{0}}}{\gamma \cosh \bar{H} t_{0}}, \\
 t_{\rm dS} &= t_{0} - \frac{1}{H_{\rm dS}} \text{Arcsinh} \left(\gamma \sinh \bar{H} t_{0}\right).
\end{align}
The relation between the original scaled closed de Sitter space and the corresponding de Sitter space can be expressed as shown in Fig.~\ref{fig2}.

To evaluate the affine length to the south pole, let us introduce the conformal time $\eta$ by
\begin{align}
 \eta = \frac{1}{\gamma} {\rm Arctan} \left( \sinh \bar{H} t \right).
\end{align}
Then $t \in (- \infty, \infty)$ corresponds to $\eta \in ( - \pi/(2 \gamma) , \pi/(2 \gamma) )$.
Now $2/\theta_{0}$ can be expressed in terms of $\eta_{0} = (1/\gamma) {\rm Arctan}(\sinh \bar{H} t_0)$ as
\begin{align}
 \frac{2}{\theta_{0}} = \frac{\gamma^2}{\bar{H}^2} \frac{1}{\cos^2 \gamma \eta_{0} }\frac{\tan \chi_{0}}{\gamma \tan \chi_{0} \tan \gamma \eta_{0} - 1}.
\end{align}
Now the positivity condition for $\theta_{0}$ can be expressed as
\begin{align}
 \tan \chi_{0} > \frac{1}{\gamma \tan \gamma \eta_{0}} \qquad \text{or} \qquad  \frac{\pi}{2} < \chi_{0} < \pi .
\end{align}
The past-outgoing null geodesics from $t = t_0, \chi = \chi_{0}$ can be expressed by $\eta(\lambda) = \eta_{0} - (\chi - \chi_{0})$. Then, the interval from $\chi = \chi_{0}$ to $\chi = \pi$ corresponds to the interval of the conformal time from $\eta = \eta_{0}$ to $\eta = \eta_{0} - \pi + \chi_{0}$.
Since the scale factor can be represented as
\begin{align}
 a(\eta) = \frac{\gamma}{\bar{H}} \frac{1}{\cos \gamma \eta},
\end{align}
the affine length to $\chi = \pi$ can be evaluated as
\begin{align}
 \Delta \lambda &= \int_{\eta_{0} - (\pi - \chi_{0})}^{\eta_{0}} a(\eta)^2  d \eta \notag\\
&= \frac{\gamma}{\bar{H}^2} \left(  \tan (\gamma \eta_{0}) - \tan( \gamma (\eta_{0} + \chi_{0} - \pi)) \right).
\end{align}
Then, we would like to show that the affine size of the universe is actually smaller than the upper bound predicted by the singularity theorem. 
Thus, let us show the positivity of the quantity
\begin{align}
 \frac{2}{\theta_{0}} - \Delta \lambda = 
\frac{2}{\theta_{0}} \frac{ f(\gamma, \pi - \chi_{0})}{\gamma \tan (\pi - \chi_{0}) \left[ 1 + \tan\bigl(\gamma(\pi - \chi_{0})\bigr)   \tan \gamma \eta_{0} \right]},\label{delta}
\end{align}
where we define a function
\begin{align}
 f(\gamma, \theta) = \gamma \tan \theta - \tan(\gamma \theta).
 \label{funcf}
\end{align}
The key properties of this function are summarized in the Appendix.

Let us first consider the case where $\pi/2 < \chi_{0} < \pi$.  In this case,  
$\tan(\pi - \chi_{0})$ and $\tan (\gamma(\pi - \chi_{0}))$, as well as $\tan \gamma \eta_{0}$, are positive and hence the denominator of the right-hand side of Eq.~\eqref{delta} is positive. By Eq.~\eqref{f1} in the Appendix, we obtain $f(\gamma, \pi - \chi_{0}) > 0$ for $0 <  \pi - \chi_{0} < \pi/2$. Thus, we can see that the right-hand side of Eq.~\eqref{delta} is positive.

Next, let us consider the case
\begin{align}
-1 + \gamma \tan \chi_{0} \tan \gamma \eta_{0} > 0.\label{thetapositive}
\end{align}
We will treat the case where the denominator of the right-hand side of Eq.~\eqref{delta} is positive or negative separately.
Assuming that the denominator is negative, then, we have
\begin{align}
 1 + \tan\bigl(\gamma( \pi - \chi_{0})\bigr) \tan(\gamma \eta_{0}) > 0.\label{demininegative}
\end{align}
By summing two inequalities \eqref{thetapositive} and \eqref{demininegative}, and by dividing it by a positive quantity $\tan(\gamma \eta_{0})$, we obtain
 \begin{align}
  \gamma \tan \chi_{0} + \tan\bigl(\gamma( \pi - \chi_{0})\bigr) = - f(\gamma, \pi - \chi_{0}) > 0.
 \end{align}
Thus, the right-hand side of Eq.~\eqref{delta} is positive in this case.

Finally, let us focus on the case where the denominator of the right-hand side of Eq.~\eqref{delta} is positive.
Since $\tan(\pi - \chi_{0})$ is negative, this can be true only when $\tan( \gamma ( \pi - \chi_{0}))$ is negative. Thus, we obtain 
\begin{align}
\frac{\pi}{2 \gamma} < \pi - \chi_{0} < \frac{\pi}{\gamma}.
\end{align}
Since we already have $\pi/2 < \pi - \chi_{0} < \pi$,
we find 
\begin{align}
 \frac{\pi}{2 \gamma} < \pi - \chi_{0} < \pi.
\end{align}
This can be satisfied only for $1/2 < \gamma ( < 1)$.
By the relation \eqref{f2} in the Appendix, we find $f(\gamma, \pi - \chi_{0}) > 0$. Thus, both the numerator and denominator of the right-hand side of Eq.~\eqref{delta} are positive. 
Thus, the inequality $\Delta \lambda < 2/\theta_{0}$ is satisfied.

\subsection{Nonsingular flat FLRW universe}
Let us consider a flat FLRW universe with a scale factor
\begin{align}
 a(t) = \frac{\bar{a} \mathrm{e}^{\bar{H} t}}{\sqrt{1+\bar{a}^2 \mathrm{e}^{2 \bar{H} t}}},\label{aexample2}
\end{align}
with positive constants $\bar{a}$ and $\bar{H}$,
that is the metric 
\begin{align}
 g_{\mu\nu}dx^{\mu} dx^{\nu} = - dt^2 + \frac{\bar{a}^2 \mathrm{e}^{2 \bar{H} t}}{1+\bar{a}^2 \mathrm{e}^{2 \bar{H} t}} (dr^2 + r^2 d \Omega^2).
 \label{nonsingflatFLRW}
\end{align}
The universe approaches the expanding flat de Sitter space $a \sim \bar{a} \mathrm{e}^{\bar{H} t}$ in the limit $t \rightarrow - \infty$,
whereas it approaches Minkowski space $a \rightarrow 1$ in the limit $t \rightarrow + \infty$.
As we will see below, the maximal extension of this spacetime is nonsingular and globally hyperbolic with a topologically $S^3$ Cauchy surface and satisfies the null convergence condition.

As discussed in Ref.~\cite{Borde:2001nh} for more general inflationary spacetime, this flat FLRW universe has incomplete null geodesics.
This does not necessarily mean the presence of singularity \cite{Vilenkin:2013rza} and as studied in Refs.~\cite{Yoshida:2018ndv, Nomura:2021lzz, Nishii:2021ylb}, the spacetime can be extended beyond the end point of incomplete null geodesics if $\lim_{t \rightarrow -\infty} \partial_t{H}/a^2$ is finite.
Actually, our example satisfies this condition because
\begin{align}
 \frac{\partial_t{H}}{a^2}  =  - \frac{2 \bar{H}^2}{1 + \bar{a}^2 \mathrm{e}^{2 \bar{H} t}}
\end{align}
is finite in the limit $t \rightarrow - \infty$. 
Then the universe can be extended beyond the past boundary of the inflationary region.
We note that this universe satisfies the null convergence condition for flat FLRW universes, $\partial_{t}H \leq 0$.

By defining new coordinates $\lambda$ and $v$ by
\begin{align}
 \lambda & = \frac{1}{\bar{H}} \text{Arcsinh} (\bar{a} \mathrm{e}^{\bar{H}t}) = \frac{1}{\bar{H}} \text{Arctanh} (a(t)),   \\
 v &= r + \eta = r  - \frac{1}{\bar{H} a(t)} + \frac{1}{\bar{H}} \text{Arctanh}(a(t)) ,
\end{align}
where $\eta = \int dt \, (1/a(t))$ is the conformal time, we can write our metric as
\begin{align}
 g_{\mu\nu} dx^{\mu} dx^{\nu} =& - 2 d\lambda dv + \tanh^2 (\bar{H} \lambda) dv^2 \notag\\
& + \frac{1}{\bar{H}^2} \Bigl( 1 + \bar{H}(v - \lambda) \tanh(\bar{H} \lambda) \Bigr)^2 d \Omega^2 .
\end{align}
Each of the $v, \theta, \phi =$ constant curves is a null geodesic and $\lambda$ corresponds to an affine parameter of such null geodesics with the normalization $d \lambda = a d t$.
The region covered by the original coordinates $t \in ( -\infty, \infty)$ corresponds to
 $ \lambda  \in \left(0 , \infty \right)$. 
The important point here is that the metric is well defined at and beyond the point $\lambda = 0$.

Actually, by defining $\tilde{\lambda} = - \lambda$ and $\tilde{v} = - {v} \in (-\infty, \infty)$,
the metric in the extended region $\lambda < 0$ can be written as
\begin{align}
 g_{\mu\nu} dx^{\mu} dx^{\nu} =& - 2 d \tilde{\lambda} d\tilde{v} + \tanh^2 (\bar{H} \tilde{\lambda}) d\tilde{v}^2 \notag\\
& + \frac{1}{\bar{H}^2} \Bigl( 1 + \bar{H}(\tilde{v} - \tilde{\lambda}) \tanh(\bar{H} \tilde{\lambda}) \Bigr)^2 d \Omega^2  .
\end{align}
Thus, the spacetime in the extended region $\lambda < 0$ is the same as the original FLRW universe in the $\lambda > 0$ region, except for flipping the future direction to the past.
Concretely, by introducing the time and radial coordinates $\tilde{t}$ and $\tilde{r}$ by
\begin{align}
\tilde{\lambda} &= \frac{1}{\bar{H}} \text{Arcsinh}(\bar{a} \mathrm{e}^{- \bar{H} \tilde{t}}), \\
 \tilde{v} &= \tilde{r} - \frac{1}{\bar{H}} \frac{\sqrt{1 + \bar{a}^2 \mathrm{e}^{- 2 \bar{H} \tilde{t}}}}{\bar{a} \mathrm{e}^{- \bar{H} \tilde{t}}} + \frac{1}{\bar{H}} \text{Arcsinh}(\bar{a} \mathrm{e}^{- \bar{H} \tilde{t}}),
\end{align}
one can write the metric in the $\lambda < 0$ region by
\begin{align}
 g_{\mu\nu}dx^{\mu} dx^{\nu} = - d\tilde{t}^2 + \frac{\bar{a}^2 \mathrm{e}^{- 2 \bar{H} \tilde{t}}}{1+\bar{a}^2 \mathrm{e}^{- 2 \bar{H} \tilde{t}}} (d\tilde{r}^2 + \tilde{r}^2 d \Omega^2).
\end{align}
This is the contracting FLRW universe obtained by flipping the time direction $t \rightarrow - t$ in Eq.~\eqref{aexample2}. 

\begin{figure}[htbp]
 \begin{center}
  \includegraphics[width=\hsize]{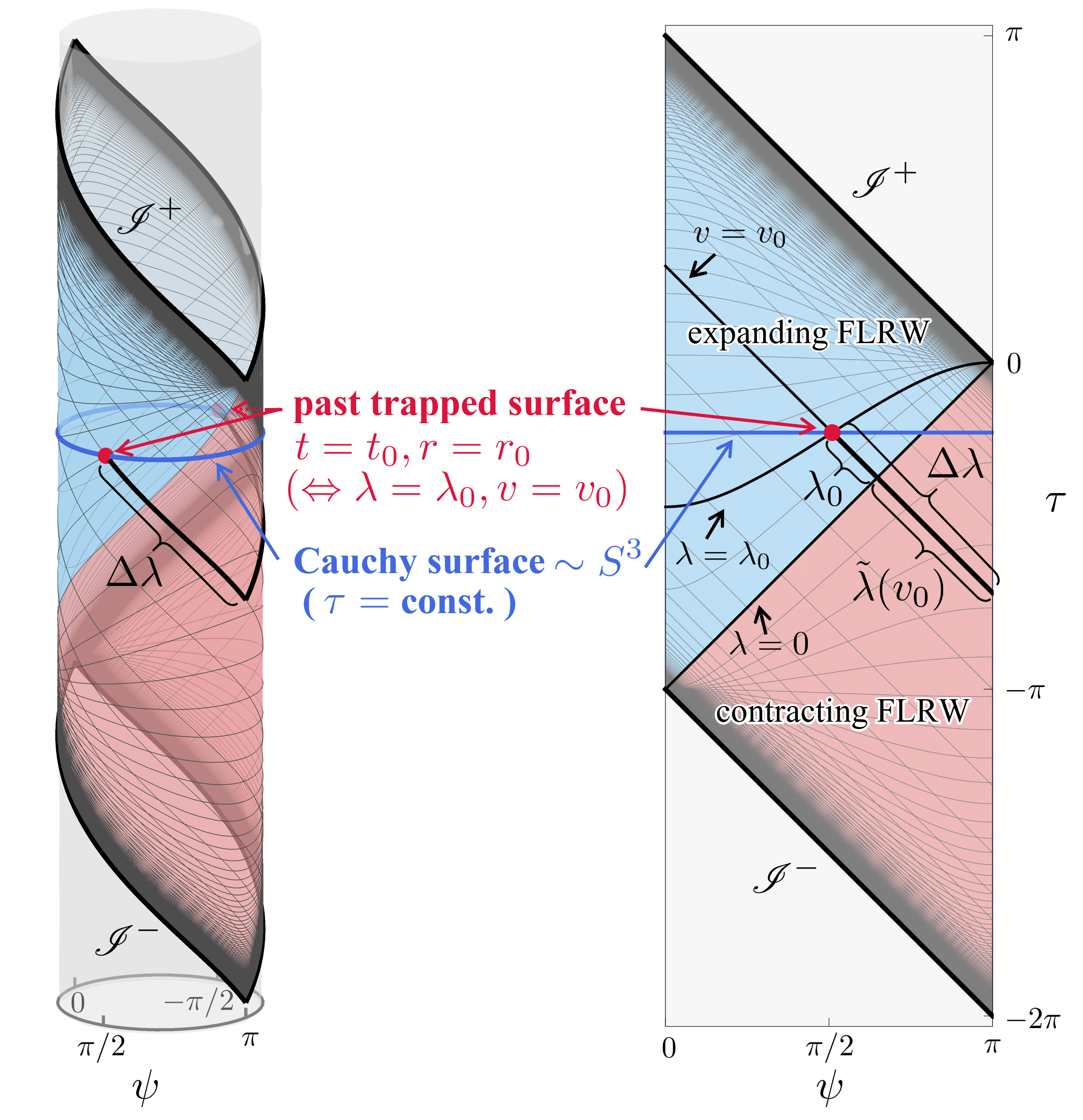}
\caption{The conformal diagram of the nonsingular FLRW universe \eqref{nonsingflatFLRW} is presented.
The left figure represents the conformal embedding of the nonsingular FLRW universe to the Einstein static universe (the whole cylinder). Here we represent the $\theta = \theta_{0}, \phi = \phi_{0}, \pi + \phi_{0}$ slice by extending the $\psi$ coordinate to $\psi \in (-\pi, \pi)$.
The right figure represents the $\tau$-$\psi$ plane with $\psi \in (0, \pi)$.
In each figure, the blue region represents the expanding flat FLRW universe $\lambda > 0$ and the red region represents the contracting flat FLRW universe $\lambda < 0$.
$\mathscr{I}^{+}$ and $\mathscr{I}^{-}$ represent future null infinity and past null infinity, respectively. The $\tau = {\rm constant}$ surface has the topology of $S^3$ and it is the Cauchy surface of the maximally extended universe. 
A past trapped surface is represented by the red points.
The black bold curves from the red points are the past-outgoing null geodesics. 
}
\label{fig3}
 \end{center}
\end{figure}

One can represent the metric as a conformal transformation of the Einstein static universe, 
\begin{align}
g_{\mu\nu}dx^{\mu} dx^{\nu} = \omega^2(\tau, \psi) \left( - d\tau^2 + d \psi^2 + \sin^2 \psi d \Omega^2 \right) ,
\end{align}
where the conformal factor is defined by
\begin{align}
 \omega^2(\tau, \psi) = \frac{1}{4} \frac{a^2(\tau, \psi)}{\cos^2 \frac{\tau + \psi}{2} \cos^2 \frac{\tau - \psi}{2}}.
\end{align}
Here the coordinate transformations from $(\lambda,v)$ coordinates are given by
\begin{align}
 \tau  = \text{Arctan} \,v - \text{Arctan}\left(v - 2\lambda + \frac{2}{\bar{H} \tanh \bar{H} \lambda}\right), \notag\\
 \psi = \text{Arctan} \,v + \text{Arctan}\left(v - 2\lambda + \frac{2}{\bar{H} \tanh \bar{H} \lambda}\right).
\end{align}
Noting that the conformal factor can be represented as
\begin{align}
 \omega^2 = \frac{4 + v^2}{16} \left[
\tanh^2 \bar{H} \lambda + \left(\frac{1}{\bar{H}} + \left(\frac{v}{2} - \lambda \right) \tanh \bar{H} \lambda \right)^2
\right]
\end{align}
and it is regular at $\lambda = 0$, which corresponds to $\tau - \psi = - \pi$. 
This universe can be conformally embedded to the subregion $- \pi < \tau + \psi < \pi$, $-3\pi < \tau - \psi < \pi,$ and $0 < \psi < \pi$ of the Einstein static universe.
Thus, the Penrose diagram can be written as Fig.~\ref{fig3}.
From the diagram, one can find that the $\tau = \text{constant}$ surface, which is expressed by the blue curves in Fig.~\ref{fig3}, is a Cauchy surface and it has the topology of $S^3$.

Let us focus on a two-sphere ${\cal T}$ defined by $t = t_{0}$ and $ r = r_{0} > (a_{0} H_{0})^{-1}$, where $a_0 = a(t_0)$ and $H_0 = H(t_0)$. 
Then the expansion of the future-ingoing null geodesic congruence is given by
\begin{align}
 \frac{2}{\theta_{0}} &= \frac{a_{0}^2 r_{0} }{r_{0} a_{0} H_{0} - 1} \notag\\
&= \frac{1}{\bar{H}} \frac{\bar{H} (v_{0} - \lambda_{0}) \sinh(\bar{H} \lambda_{0}) \cosh(\bar{H} \lambda_{0}) + \cosh^2(\bar{H} \lambda_{0})}{\bar{H} (v_{0} - \lambda_{0}) - \sinh(\bar{H} \lambda_{0}) \cosh(\bar{H} \lambda_{0})}.
\end{align}
Here we introduced $\lambda_{0}$ and $v_{0}$ as the value of $\lambda$ and $v$ on the trapped surface ${\cal T}$. 

On the other hand, the outward affine length to the origin of the contracting FLRW universe can be evaluated as
\begin{align}
 \Delta \lambda  = \lambda_{0} + \tilde{\lambda}(v_{0}),
\end{align} 
where $\tilde{\lambda}(v_{0})$ is the affine length from the origin of contracting universe to the boundary $\lambda = 0$.
It can be defined as the positive root of the equation
\begin{align}
  \bar{H} v_{0} + \bar{H} \tilde{\lambda}  = \frac{1}{\tanh (\bar{H} \tilde{\lambda})}.
\end{align}
Note that $\tilde{\lambda}(v_{0})$ is independent from $\lambda_{0}$.

What we want to confirm is the non-negativity of the quantity,
\begin{align}
\delta(\lambda_{0}, v_{0}) =  \frac{2}{\theta_{0}} - \Delta \lambda.
\end{align}
One can check that $\lambda_{0}$ derivative of $\delta(\lambda_{0}, v_{0})$ is always non-negative,
\begin{align}
&\partial_{\lambda_{0}} \delta(\lambda_{0}, v_{0}) \notag\\
&\quad = 2 \left( \frac{\cosh \bar{H} \lambda_0 + \bar{H} (v_{0} - \lambda_{0}) \sinh \bar{H}\lambda_{0} }{-\bar{H} (v_{0} - \lambda_{0}) + \sinh(\bar{H} \lambda_{0}) \cosh(\bar{H} \lambda_{0})}\right)^2
\geq 0.
\end{align}
Then $\delta(\lambda_{0}, v_{0}) \geq \delta(0, v_{0})$ for any $\lambda_{0} \geq 0$.
We can express $\delta(0, v_{0})$ as
\begin{align}
 \delta(0, v_{0}) = \frac{1}{\bar{H}^2 v_{0}} - \tilde{\lambda}(v_{0})
\end{align}
and the positivity of $\delta(0, v_{0})$ can be shown as follows:
$\bar{H} \tilde{\lambda}(v_{0})$ is defined as the value of $x>0$ at the intersection of 
\begin{align}
 y = x + \bar{H} v_{0},\quad   \text{and} \quad  y = \frac{1}{\tanh x}.
\end{align}
On the other hand, $1/(\bar{H} v_{0})$ is the value of $x>0$ at the intersection of 
\begin{align}
 y = x + \bar{H} v_{0},\quad  \text{and} \quad y = x + \frac{1}{x}.
\end{align}
Since
\begin{align}
  \frac{1}{\tanh x} < x + \frac{1}{x} \qquad (x > 0),
\end{align}
we obtain $1/(\bar{H} v_{0}) > \bar{H} \tilde{\lambda}(v_{0})$ and hence we obtain $\delta(\lambda_{0}, v_{0}) \geq \delta(0, v_{0}) > 0$. 

\section{Constraint from Observations}
\label{sec:observation}
In this section, we discuss the implications of the statement on the size of a nonsingular universe for our Universe. 
Let us assume that our Universe satisfies global hyperbolicity and the null convergence condition and that our Universe is past null geodesically complete.
Then, if there is a past trapped surface, the Cauchy surface must be compact and the outward affine size of our Universe has an upper bound as stated in Sec.~\ref{sec:singularitytheorem}.
Since current cosmological observations are consistent with the FLRW universe with flat spatial curvature, $k=0$, we describe our Universe as such. 
As mentioned in Sec.~\ref{sec:FLRW}, then it is expected that a two-sphere with a sufficiently large radius centered on us is a past trapped surface.
In fact, recent observations of the scale of baryon acoustic oscillations (BAOs) have successfully measured the expansion rate at high redshift so that we can find the existence of a past trapped surface. We employ the analysis results of Ref.~\cite{duMasdesBourboux:2020pck} where the BAO measurements at the redshift $z = 2.33$ are presented. 
The obtained Hubble radius $D_H(z) = 1 / H(z)$ 
and comoving distance $D_M(z)$ at that redshift are $D_H(z=2.33)/{r_d} = 8.99$ and $D_M(z=2.33)/{r_d} = 37.5$, respectively, where $r_d = 147.3$ Mpc is the sound horizon at the decoupling epoch.
Note that these values are derived directly from observations and do not rely on any specific cosmological model such as the cosmological constant and cold dark matter model. 
Then we have $(aH)^{-1}/{r_d} = (1+z) D_H/{r_d} = 29.9$ at $z = 2.33$, which is less than $D_M(z=2.33)/{r_d}$. Thus, the inequality \eqref{trapped} for $k=0$ is satisfied, which means that the two-sphere at that distance is a past trapped surface.
Therefore, under the assumptions of global hyperbolicity, null convergence, and the absence of initial singularity, by using Eq.~\eqref{lambdaineq} we can give the upper bound on the outward affine size of our Universe from the trapped surface at $z=2.33$ as 
\begin{align}
\Delta \lambda \leq \left. \frac{a^2}{aH - D_M^{-1}} \right|_{z=2.33} = 1.97 \times 10^3 \, \mathrm{Mpc}.
\end{align}
Note that we use the affine parameter defined by $d\lambda = a dt$ and the normalization of $a$ is chosen so that $a = 1$ at the current time. 
If we use alternative definition $ d \lambda = (a(t)/a(t_{0})) dt $, where $t_{0}$ is the time at the trapped surface, the inequality can be rephrased as
\begin{align}
 \Delta \lambda \leq 6.57 \times 10^3 \, \mathrm{Mpc}.
\end{align}
With this normalization, the affine length represents the spatial distance traveled by light if the outside of the trapped surface is replaced by Minkowski space.

Furthermore, the statement on de Sitter space as the largest space mentioned in Sec.~\ref{subsec:dSlargest} implies that the affine size of our Universe must be smaller than that of a universe in which the region prior to $z=2.33$ is replaced by de Sitter space with the Hubble rate $H(z=2.33) = 226$ km/sec/Mpc. 
It is widely expected that, as going back to the past, the description of our Universe turns from the big bang model into another scenario such as inflation or a bouncing universe, and thus the initial singularity of big bang cosmology would be absent. The result here then states that such a scenario without initial singularity must have a finite size less than or equal to the above bound if the global hyperbolicity and null convergence condition continue to hold.

\section{Summary and Discussion}
In this paper, we revisited Penrose's first singularity theorem and clarified the implications of the theorem for nonsingular universes, aiming to understand a general property of spacetimes realized in a class of gravitational theories that is consistent with the null convergence condition. As a direct consequence of the singularity theorem, a globally hyperbolic expanding universe consistent with the null convergence condition must have a compact Cauchy surface, of which the universal covering must be topologically $S^3$. Then we found that the outward affine size of the universe from a past trapped surface, defined through the affine parameter of the past-outgoing null geodesic in Sec.~\ref{sec:2B}, is bounded above by a value determined by the initial expansion on the trapped surface. Applying this result to homogeneous and isotropic universes, we found that the upper bound corresponds to the outward affine size of the de Sitter universe and hence we found that a nonsingular FLRW universe is always smaller than de Sitter space. We provided two examples of nonsingular universes that are consistent with the null convergence condition and confirm directly that the outward affine size of each universe is smaller than that of the de Sitter universe. Then we applied our result to a past trapped surface at the redshift $z = 2.33$ in the realistic Universe, which is directly confirmed by BAO measurements. We found that the outward affine size of the Universe measured from the trapped surface must be smaller than $6.57 \times 10^3 ~\text{Mpc}$ with the normalization $d \lambda = (a/a(z = 2.33)) dt$, as long as our Universe is nonsingular, globally hyperbolic, and consistent with the null convergence condition. 
Note that we do not assume the Einstein equations, as is the singularity theorem, and hence our results hold even when the dynamics of gravity is modified due to the effect from UV physics, as long as the null convergence condition holds.

We would like to emphasize that the flat FLRW universe that we investigated in Sec.~\ref{sec:example} itself is a good analytic example of discussing the nature of a nonsingular spacetime because it has the following properties: it is nonsingular, globally hyperbolic, and consistent with the null convergence condition, possesses the symmetry of flat FLRW spacetime, and reduces to Minkowski space near future and past null infinity. 

In the present paper, we focused only on the universe consistent with the null convergence condition, which is equivalent to the null energy condition when the Einstein equations hold. It should be interesting to extend our discussion to more general situations where the null energy condition is violated, for example, using a kind of averaged null energy condition \cite{Wald:1991xn,Graham:2007va} instead of the null energy condition.

\begin{acknowledgments}
D.Y.~thanks Ken-ichi Nakao for fruitful discussions.
K.N.~was supported by Grant-in-Aid for JSPS Research Fellowship and JSPS KAKENHI Grant No.JP21J20600.
D.Y.~was supported by JSPS KAKENHI Grants No.JP20K14469 and No.JP21H05189.
\end{acknowledgments}

\appendix
\section{PROPERTIES OF THE FUNCTION $f(\gamma, \theta)$}
In this appendix, we summarize some properties of the function 
\begin{align}
 f(\gamma, \theta) = \gamma \tan \theta - \tan \gamma \theta,
\end{align}
which appeared in Sec.~\ref{sec:example}.

The first property we used there is
\begin{align}
  f(\gamma, \theta) > 0 \qquad \text{for }~ 0 < \theta < \frac{\pi}{2},\, 0 < \gamma <1. \label{f1}
\end{align}
This inequality can be shown by evaluating the $\gamma$ derivative of $f/\gamma$,
\begin{align}
 \partial_{\gamma} \left( \frac{f(\gamma, \theta)}{\gamma}  \right)
&=
- \frac{2 \gamma \theta - \sin (2 \gamma \theta)}{2 \gamma^2 \cos^2 (\gamma \theta)}  < 0.
\end{align}
Since $f(\gamma, \theta)/\gamma$ is continuous in $0 < \gamma \leq 1$ and $f(\gamma, \theta)/\gamma |_{\gamma = 1} = 0$, $f(\gamma, \theta) > 0$ holds for $0 < \gamma < 1$.

The other property that we used in the main section is
\begin{align}
 f(\gamma, \theta) > 0 \qquad \text{for }~ \frac{\pi}{2 \gamma} < \theta < \pi,\, \frac{1}{2} < \gamma < 1.\label{f2}
\end{align}
This inequality can be shown by evaluating the $\theta$ derivative of $f$,
\begin{align}
 \partial_{\theta}f(\theta, \gamma) = \frac{\gamma}{\cos^2 \theta} - \frac{\gamma}{\cos^2 \gamma \theta}.
\end{align}
Since $\cos^2 \theta > \cos^2 \gamma \theta$ for $\pi/(2 \gamma) < \theta < \pi$ and $ 1/2 < \gamma < 1$, we obtain $\partial_{\theta} f < 0$.
In addition, $f(\gamma, \theta)$ is continuous in $\pi/(2\gamma) < \theta \leq \pi$ and $f(\gamma, \pi) = - \tan(\gamma \pi) > 0$, so we can obtain $f(\gamma, \theta) > 0$ in this region.

\bibliography{ref}
\end{document}